\documentstyle[12pt]{article}
\begin{document}
\bibliographystyle{unsrt}
\baselineskip=20pt
\centerline{\large \bf CAN WE SEE THE HADRON-QUARK TRANSITION}
\centerline{\large \bf  HAPPENING IN NEUTRON STARS?}
\vspace{0.25in}
\begin{center}
Fr\'{e}d\'{e}rique Grassi\\
{\em Instituto de F\'{\i}sica, Universidade de S\~ao Paulo\\
C.P. 66318, 05315-970 S\~ao Paulo-SP Brazil}\\
and\\
{\em Instituto de F\'{\i}sica, Universidade Estadual de Campinas\\
C.P. 1170, 13083-970 Campinas-SP Brazil}\\
\vspace{0.5in}
{\bf A B S T R A C T}\\
\end{center}
In order to actually see the hadron-quark transition happening in a neutron 
star, we point out and study two static conditions (the transition
hadronic density must be lower than the neutron star maximum hadronic density;
the neutron star mass at the transition hadronic density must be in the observed
range $\approx 1.4 M_{\odot}$) and one dynamical condition (nucleation must
occur during the star lifetime).
We find that the mini-colapse acompanying the transition from metastable 
hadronic matter to quark matter may be relevant to explain macro-glitches and 
gamma ray bursts, but that the mecanism increasing the
star density must be relatively fast, e.g. accretion but not slowing down.
This rules out a scenario for gamma ray bursts proposed recently.

{\em subject heading}:dense matter - elementary particles - gamma rays:
bursts - stars: neutron

\newpage

\noindent{\bf 1. Introduction}

A first order transition may be trigerred in a neutron star
whose central density is increasing. Such an increase
 may come from the fact that the star 
is slowing down, accreting
matter, or cooling.
For a second order phase transition or a first order transition occuring
in equilibrium,
 all the macroscopic quantities of the
star (mass, radius, momentum of inertia) 
vary continuously (the only discontinuity being -possibly- in the central 
density). 
For a first order transition occuring out of equilibrium, with the appearance of
a metastable, supercompressed neutron phase,
the star undergoes a mini-colapse when
it
goes from one configuration with
a  metastable neutron core, a mass $M_1$ and radius $R_1$,
 to another configuration with mass, $M_2$, and radius $R_2$, consisting of
a core made of the new phase, surrounded by normal
neutron matter, 
(all this occuring at constant baryonic number). In the mini-colapse,
a large amount of energy may be released in a
short time,
just how big is this amount of energy, depends on the equation of
state inside the neutron star, how quickly the new phase nucleates, etc, so
it has to be determined for the particular type of new phase considered : pion
matter, quark matter, kaon condensate...
A more familiar example of a metastable system is a gas undergoing a first order
phase transition to its liquid if compressed to a volume $V_c$ at fixed T. If 
it is slowly compressed at fixed T, it may stay in the vapor state even for
$V<V_c$. Other examples are liquid water freed of dissolved air which may
be heated above boiling temperature without the formation of vapor
and cooled below freezing temperature without solidification.

Various papers have tried to explore the possibility of a mini-colapse during 
a first order transition.
Migdal \& al. (1979)
first
studied the dynamical effects of the neutron-pion transition in
dense stars, and showed that $10^{53}$ erg of energy could be emitted in 
$10^{-4}$ s, this lead them to suggest that this transition could provide an
alternative explaination to the supernova phenomena (see also Berezin et
 al. 1982). 
This calculation was taken up by Haensel \& Pr\'{o}szy\'{n}ski (1980,1982)
 with a more realistic pion equation of state. These authors concluded
that the amount of energy released would probably be smaller than previously  
predicted, $\sim 10^{48}$ erg, but that
because of the change in the momentum of inertia (0.1\%), the pion transition 
might cause observable macro-glitches. 
It was also  proposed that a pion transition might be at the origin of the
March-5-1979 $\gamma$ transient (Ramaty et al. 1980, Ellison \& Kazanas
1983).
Finally, details of the
actual relaxation from a one phase star
 to a two phase star via oscillations of the pion core surface, 
were worked out by K\"{a}mpfer (1981)
and
 Diaz-Aloso (1983)
for simplified  equations of state. However,
 measurements of the  coefficient g' (Carey et al. 1984)
of the short-range spin-isospin
repulsion seem to exclude the possibility that the pion transition 
 happen in stable neutron stars (Baym \& Campbell 1979).
In principle, another type of condensate, the kaon condensate, may exist with
effects similar to a pion condensate (Brown \& Bethe 1994)
 but
the details of kaon condensation are 
very sensitive to the kaon-dense matter interactions (Pandharipande
 et al. 1995).
Finally, recently it was suggested that that the hadron-quark transition
occuring in  slowing down neutron stars might be responsible for Gamma Ray 
Bursts (Ma \& Xie 1996).

In this paper, we concentrate on the following question: can we hope to see 
the hadron-quark transition happen in a neutron star (via the mini-colapse
manifestations) ? The question can be divided in three conditions
that must be satisfied to get a positive answer.
First, as usual, the transition density must be smaller than the maximum
central density of a pure neutron star. Second, the mass of the neutron star
at the transition should be around
$1.4 M_{\odot}$ 
since this is the typical mass of the neutron stars observed so far
(the errors bars  on measured masses are big, 
but taking them into account only widens the interval of permited
parameters for the quark matter equation of state).
If the transition is predicted to occur in stars outside that mass range,
since such stars are not observed, we would not see the 
transition happening.
Finally,
normally one expects that at $n_{H\,t}$, the neutron and quark phases coexist.
However, due to surface effects, the neutron phase may survive for a range of 
densities $> n_{H\,t}$ (it is said to be metastable) during a time
called its lifetime or nucleation time, after that quark matter droplets 
appear. 
The nucleation time decreases with increasing 
density and
 becomes smaller
than the estimated neutron star lifetime at some hadronic
density $n_{H\,'}>n_{H\,t}$. The third condition will therefore be that
the mecanism which increases the central density, starting say from
$n_{H\,t}$, can bring it to $n_{H\,'}$ in less than the star lifetime
so that the transition will actually occur.
The first two conditions correspond to a static problem: we only need to
solve the Tolman-Oppenheimer-Volkoff (TOV) equations.
The third condition corresponds to a dynamical problem,
the nucleation of quark droplets in cold hadronic matter.

\noindent{\bf 2. The static problem}

In order to solve the equations of hydrostatic equilibrium, we need to
specify the equation of state.
The equation of state of quark matter is taken to be the M.I.T. one (Baym \& 
Chin 1976),
i.e.
$
\epsilon=An^{4/3}+B
$,
$
p=An^{4/3}/3-B
$.
The values B=55 MeV fm$^{-3}$, $\alpha=2.2$, give a reasonable fit to
observed hadronic masses in the original MIT analysis, 
however when dealing with dense stars, one has no good clue
as to what the value of these parameters should be. We will therefore work to 
zeroth order in $\alpha$, so that for two flavor electrically neutral quark 
matter,
$A=3/4 \pi^{2/3} (1+2^{4/3})$, and we will keep B as a parameter. 
By  doing so, even though the exact equation of state of quark matter
is unknown, we can enumerate and study various possible scenarios for the 
transition, which arise according to the value of B. 
(We consider two flavors because when the hadrons ``melt''
at the transition, mostly u and d quarks are present for the 
hadronic equations of state that we use; s quarks appear later when the bubble
is formed already via weak 
interactions. To make the transition directly from mostly neutron matter to
u-d-s matter would require several simultaneous weak interactions, which is
unlikely).

The equation of state for normal neutron stars used, consists of, as usually
done, a matching of 
 the Feynman-Metropolis-Teller (1949),
Baym-Pethick-Sutherland (1971)
and
Baym-Bethe-Pethick (1971)
 equations of states. In addition,
for the star center, 
two equations are considered : a soft one, the model "VN" of Bethe-Johnson
(1974)
denoted as BJVN,
(softer equations of state predict too low maximum masses for neutron
 stars) and a stiff one, the Walecka (1974) equation of state;
the real equation of 
state of hadron matter should lie between these two extremes.

In order to determine the density for the
 hadron-quark transition, we plot the pressure p as a function of
the Gibbs energy or baryonic chemical potential
($\equiv (\epsilon+p)/n_B$) for various values of B and for
the BJVN equation of state
 or the Walecka equation of state,
then look for intersections, as they signal phase transitions\footnote{
We study here the standard case of neutron and quark matters coexisting
for a single pressure and not a range of pressures as suggested by Glendenning
(1992),
first because our formalism is easier to construct in that case, second because
it is not established which of these two possibilities is correct (see Heiselberg
et al. 1993)}. 

We can then proceed  to solve the TOV equations . For a neutron star with a BJVN equation of state
and no quarks
inside, the maximum mass is M$_{max}$=1.76$M_\odot$ at a radius R=9.2 km
and a central density n$_c$=1.49 fm$^{-3}$. In the case of a 
star with a Walecka 
equation of state and no quarks
inside, M$_{max}$=2.59$M_\odot$ at R=12.4 km
and n$_c$=0.76 fm$^{-3}$.
In Table 1, we show the mass M$_t$ of the neutron star when the transition 
occurs, (also shown are its momentum of inertia $I_t$ and radius $R_t$). 
For a BJVN equation of state, except for a finely tuned B$\sim 56$ MeV 
fm$^{-3}$, condition 1 and/or 2 is not satisfied.
On the other side, for a Walecka equation of state, for a large
interval of B's, conditions 1 and 2 are satisfied.
Therefore if a hadron-quark transition occurs in a neutron star with an
equation of state somewhat stiffer than the BJVN one, there is hope
that we could see some effects associated with it. 
In  the remaining of this section, we  
concentrate on such a prototype, the Walecka equation of state.

In figure 1, we show the M(R) curve around the transition, for a neutron star
with a Walecka type equation of state. 
As mentioned in the introduction,
two scenarios can occur:
1) the phase transtion occurs with equilibrium of $p$ and $G$ as the
central density increases, and the star passes smoothly from the neutron 
branch (dashed line) to the quark core branch (solid branch) at the 
bifurcation point 
2) as its central density increases, the star continues evolving along the 
neutron star branch past the bifurcation point,
when
the star actually makes a transition to a quark core configuration, 
a certain quantity of energy (depending on the amount of metastability)
will be emitted in a short time.   This energy can be used to heat the star 
core, produce vibrations, etc, so it is not emitted all at once. The upper
bound for the energy released during the collapse is 
$
E=M_1-M_2
$
A lower bound for the duration of release of  energy can be 
estimated by computing the free fall time
$
\tau_{ff}=\pi\,\sqrt{R_1^3/(8GM_1)}
$.
This does not take into account all the physics of the transition 
(e.g. matter does not fall freely but is slowed by pressure gradients, the 
matter at the periphery of the metastable core probably does not make the
transition at the same time than the one at the center, etc) ;
in the case of the pion transition, its use is supported by results from numerical 
computations.

In table 1, $\tau_{ff}$, E, $\Delta I$ and $\Delta R$ are shown,
 for comparison,  for various degrees of metastability. These quantities are 
of the same order of magnitude as in
 the pion transition (Haensel \& Pr\'oszy\'nski 1980,1982)
so one may conclude also, that
the amount  of energy relased and the time for it, makes the quark transition
 relevant to explain macroglitches or
 $\gamma$ transients such as the March-5-1979 one, but not supernovae.
(Note that not the whole released gravitational energy is expected to be 
transformed in $\gamma$ rays).

\noindent{\bf 3. The dynamical problem}

We now turn to the calcution of the nucleation time. To do that, we
adapt to quark matter
the formalism developed by Lifshitz \& Kagan (1972)
and Haensel \& Schaeffer (1982).

The nucleation of quark bubbles  occurs as follows.
First the hadronic density increases above the transition hadronic density.
The formation of a quark droplet of negligeable size occurs instantaneously.
It may decay back to the hadronic phase or increase its size  by quantum 
fluctuation. This process depends on two parameters, the baryonic content of
the quark droplet, A, and its baryonic density, $n_q$. To simplify,
one considers fluctuations with fixed $n_q$ and varying A.

We write the hamiltonian describing the behavior of A, as
\begin{equation}
H(A,n_q)=K(A,n_q)+W(A,n_q)
\end{equation}
Considering the surface of the bubble as a discontinuity
and writing the equality of baryonic flux on both sides of the
surface as well as the continuity equation inside and outside the bubble, 
we get the velocity field
\begin{eqnarray}
v(r)= & 0 & if\,\,\,r<R\\
      & -\frac{\dot{R} R^2}{r^2} \left( \frac{n_q}{n_H}-1 \right)
& if\,\,\, r >R
\nonumber 
\end{eqnarray}
hence the kinetic energy reads
\begin{equation}
K(A,n_q)=M(A,n_q) \dot{A}^2
\end{equation}
with
$ M(A,n_q)=\frac{1}{2} 3 m \frac{n_H}{n_q} \left( \frac{n_q}{n_H}-1 \right)^2
\frac{1}{9} \left( \frac{4 \pi}{3} n_q \right)^{-2/3} A^{-1/3}$.

Then we turn to the potential term. Since we are going to suppose nucleation
 at constant temperature and pressure, the work needed to create a bubble
is just the increase in its Gibbs free energy
\begin{equation}
W(A,n_q)=\Delta G(A,n_q) \equiv G_q-G_H
=\mu_q N_q-\mu_H N_H + 4 \pi R^2 {\cal S} + 8 \pi R \gamma
\end{equation}
The first two terms correspond respectively to bulk quark and hadronic matter,
the next terms, the surface term and the curvature term, arise
because the quark matter-hadronic matter interface is not plane.
But since we consider u and d quarks, i.e. massless quarks, the
surface term vanishes (Berger \& Jaffe 1987,1991).
Using $\mu_u \equiv \mu_q= 2^{-1/3} \mu_d$ and the fact that the baryonic
number in quarks equal that in hadrons, i.e. A, we get
\begin{equation}
\Delta G(A,n_q)= g_v(n_q) A + g_l(n_q) A^{1/3}
\end{equation}
with $g_v=(1+2^{4/3})\mu_q-\mu_H$ and $g_l=8 \pi^{2/3} (3/4\pi)^{1/3}
1/\mu_q \gamma$, with $\gamma=(1+2^{2/3})/8\pi^2 \mu_q^2$ (Madsen 1993)
For comparison, in previous studies (Haensel \& Schaeffer 1982)
pions have a surface term and no curvature.

From this form, we see that $\Delta G$ starts from 0 for A=0, increases
to a maximum, $\Delta G_{max}=1/[3 \sqrt{3} (-g_v)]+g_l/[\sqrt{3} (-g_v)]$,
at $A_c=(-3 g_v)^{-3/2}$ then decreases, crossing 0 for $A_0=(-g_l/g_v)^{3/2}$ 
and going  to $-\infty$ for infinite A's.
Therefore quark bubbles with
$A \geq A_0$ increase without bond because this decreases $\Delta G$.
Such bubbles can be created by tunneling effect from $A=0$ to $A=A_0$, 
(which have equal values of $\Delta G$). Using the WKB approximation for the
probability,
the nucleation time reads
\begin{equation}
t_{nucl}=t_{IF} \times \exp W
\end{equation}
with $t_{IF}\sim$ 
1 fm and 
$
W=\Delta G_{max}/\tilde{T}
$
with
$
\tilde{T}=\frac{8}{3 \sqrt{3}}  \left[ \sqrt{\frac{-g_v 2 n_q}{3 m n_H}
\left(\frac{4 \pi}{3} n_q \right)^{1/3} A_0^{-1/3} \frac{n_H}{n_q-n_H}}
\right]
$, $n_q$ and $n_H$ are related one to the other by the equality of pressures
$p_q=p_H$. We are therefore ready to calculate the nucleation times
for various $n_H>N_{H t}$.

\noindent{\bf 4. Astrophysical consequences}

In table 2, we show the results of such a calculation.
For B=56 MeV fm$^{-3}$ and a BJVN equation of state, 
the transition occurs for  $n_{H t}=1.46$ fm$^{-3}$.
One sees that when $n_H$ goes from 1.62 fm$^{-3}$ to
1.63 fm$^{-3}$, the nucleation time falls from $10^{11}$ yr to $10^5$ yr, i.e.
at that density nucleation happens during
the star lifetime.
(Based on spin-down, one gets a typical value for this of $10^7$ yr, but higher 
values are possible. We take $10^7$ to be conservative in what follows).
The question that arises then, is whether
there exist astrophysical mecanisms able to incrase the central density of
a neutron star, from $n_{H t}$ or less, to $n_H\approx$ 1.63 fm$^{-3}$ 
in less than the star lifetime. A similar observation holds for the
Walecka equation of state.

Ma \& Xie (1996) estimated in their paper, that due to slowing down, the increase
in central energy density during the star lifetime is 
$\Delta \rho_{centr}/\rho_{centr} \simeq 10^{-3}$. 
In term of hadronic density,
starting from $n_{H t}$, this corresponds to an increase of $10^{-3}$, i.e. 
for practical purposes the nucleation time remains infinite 
(see table 2)and the transition 
does not happen during the star lifetime. This rules out their explaination
of gamma ray bursts.

A more optimistic scenario arises when considering an accreting neutron star.
For a BJVN star, $dM/dn_{centr\,\,\,\mid n_{H t}} \sim 10^{-2}$. Using the typical accretion
rate of $10^{-9} M_{\odot} yr^{-1}$, we see that a star goes from
$n_{H t}$ to $n_H=1.63$ fm$^{-3}$ in $\sim 10^{6}$ yr, i.e. less than
its lifetime. In other words, an accreting neutron star may be able to make
the hadron-quark transition during its life. A similar argument holds
for a Walecka
equation of state. However since accretion is a complicated 
 mecanism and its description is controversial, it would be interesting to
 check the precise behavior of central density for
various scenarios to be more afirmative.

Estimates of $\Delta n/n$ due to cooling and subsequent nucleation times are
underway.

Note that 
even if we could not find an astrophysical mecanism permiting neutron stars
to make the  hadron-quark transition 
during their lifetime, this would not rule out the possible existence
of quark core neutron stars, simply they may have to be born like that.
Also we are only making estimates of quantum nucleation,
early in its life, a star could undergo thermal nucleation
(but this phase does not last very long).

\noindent{\bf 5. Conclusion}

In summary, we studied the possibility of 
actually seeing the
transition from hadron matter
to quark matter occuring in neutron stars.

Since the dense matter equation of state is unknown and to be rather general,
we used the MIT bag equation of state for quark matter 
with varying values of B 
and two different hadronic equations of state, namely a soft and a stiff one.
We first impose the condition that quark core neutron stars be stable, i.e.
that $n_{H t} < n_{max}^{n*}$, this restricts the B-space available
[first condition].
Imposing  that the mass of the star at the transition is above 1.4 M$_{\odot}$
(if below, all observed
neutron star would already have a quark core)
further restricts the B-space that is interesting for us [second 
condition].
Next we calculated the nucleation time as a function of the metastable
hadronic matter density. We showed that nucleation may occur
during a star lifetime in a process where central density increases
fast (e.g. not slowing down but accretion, which invalidates the model by Ma \& Xie
(1996)
 [condition 3]. 
(Consideration of the inclusion of the strong coupling constant, density
fluctuations, etc, will be presented elsewhere but does not change
noticeably the results presented here.)

This opens the possibility of detecting some effects of the quark transition.
It was not the case
in previous studies for the pion transition (namely condition 1 or 2 was not 
satisfied).
A first estimate of the energy released and time for
release shows that the minicolapse metastable matter-quark matter, may be 
relevant to explain macroglitches or gamma ray busrts, but not supernovae.

This work was partially supported by FAPESP (proc. 95/4635-0),
MCT/FINEP/CNPq (PRONEX) under contract 41.96.0886.00 and CNPq
(proc. 300054/92-0).

\newpage

{\bf References}

G.Baym, A.Bethe, and C.Pethick 1971,
Nucl.Phys.A 175, 225.

G.Baym and D.Campbell 1979, in
'Mesons in nuclei', Eds. M.Rho and D.H.Wilkinson, North-Holland
  Publ.Co., Vol.III p.1031

G.Baym and S.A.Chin 1976,
Phys.Lett.B 62, 241.

G.Baym, C.Pethick, and P.Sutherland 1971,
Astrophys. J. 170, 299.

Yu.A.Berezin, O.E.Dmitrieva, and N.N.Yanenko 1982,
Sov.Astron.Lett. 8, 43

M.Berger and R.Jaffe 1987,
Phys. Rev. C 35,213

M.Berger and R.Jaffe 1991,
Phys. Rev. C 44, 566 (E)

H.A.Bethe and M.B.Jonhson 1974,
Nucl. Phys. A 230, 1.

Brown G.E. and Bethe H.A. 1994,
Ap.J. 423, 659

T.A.Carey and al. 1984
Phys.Rev.Lett. 53, 144

J.Diaz Alonso 1983,
Astron.Astrophys. 125, 287

D.C.Ellison and D.Kazanas 1983,
Astron. Astrophys. 128, 102.

R.P.Feynman, N.Metropolis, and E.Teller 1949,
Phys. Rev. 75, 1561.

N.K.Glendenning 1992, 
Phys. Rev. D 46, 1274.

P.Haensel and M.Pr\'{o}szy\'{n}ski 1980,
Phys.lett.B 96, 233.

P.Haensel and M.Pr\'{o}szy\'{n}ski 1982,
Astrophys. J. 258, 258.

P.Haensel and R.Schaeffer 1981,
Nucl. Phys.A 381, 519.

H.Heiselberg, C.J.Pethick and E.F.Satubo 1993,
Phys. Rev. Lett. 70, 1355.

B.K\"{a}mpfer 1981,
Phys.Lett.B 101,366.

I.M.Lifshitz and Yu.Kagan 1972,
JETP 35, 206.

Ma~F. and Xie B.,
Ap.J.Lett.462.

Madsen J. 1993,
Phys. Rev.Lett. 70, 391.

A.B.Migdal, A.I.Chernoutsian, and I.N.Mishustin 1979,
Phys.Lett.B 83, 158

Pandharipande~V.R, Pethick~C.J. . and Thorsson V. 1995,
Phys.Rev.Lett. 75, 4567.

R.Ramaty et al. 1980, Nature 122, 287

J.D.Walecka 1974,
Ann.Phys. 83, 491.

\newpage
\noindent {\bf Figure captions}\\ \\
\noindent Fig.1 : Behaviour of the massin M$_{\odot}$ as a function of radius
in km.
The dashed line corresponds to a (Walecka) neutron star without transition,
solid line to quark core configurations.
\\ \\ \\
\noindent {\bf Table caption}\\ \\
\noindent Table 1 : Various quantities of interest at the transition (see text for notations)
in the case of neutron matter with a soft equation of state (Bethe-Jonhson)
or a stiff equation of state (Walecka). The first set of E, $\Delta R$, $\Delta I$,
$\tau_{ff}$, corresponds to a transition occuring in metastable neutron matter of density
$n_H^t+5$\%, and the second set
corresponds to a density $n_H^t+15$\%. \\
\\
\noindent Table 2: For various values of the density of the metastable phase, 
n$_{H t}$,
calcuations of $A_0$ and the exponent W' in $\tau_{nucl}=10^{-30}\times 10^{W'}$
yr. The MIT bag equation with B=56 Mev fm$^{-3}$ (resp. 73)
and BJVN (resp. Walecka) equation of
state are used in the first (resp. last) three rows.

\newpage

\textwidth 22.cm
\hoffset=-3.8cm
\small{
\begin{tabular}{|c|l|l|l|l|l|l||l|l|l|l||l|l|l|l|}
\hline
  & B MeV & n$_{H t}$ & n$_{q t}$ & M$_t$ & I$_t$ 10$^{45}$ & R$_t$ & E 10$^{50}$ &
$\Delta$ R & $\Delta$ I & $\tau_{ff}$ &
E 10$^{50}$ &
$\Delta$ R & $\Delta$ I & $\tau_{ff}$ \\
  & fm$^{-3}$ & fm$^{-3}$ & fm$^{-3}$ & M$_{\odot}$ & g cm$^{-2}$ & km &
erg & & & 10$^{-4}$ s& erg & & & 10$^{-4}$ s\\ \hline
soft & 56. & 1.46 & 1.85 & 1.7582 & 1.3 & 9.2 & - & - & - & - & - & - & -
& - \\
     & 66. & 0.23 & 0.29 & 0.3095 & 0.2 & 13.1 &- & - & - & - & - & - & -
& - \\
     &     & 1.50 & 1.94 & unst.  & -   & -    &- & - & - & - & - & - & -
& - \\
     & 68. & 1.50 & 1.95 & unst.  & -   & -    &- & - & - & - & - & - & -
& - \\ \hline
stiff& 56. & -    &  -   & -      & -   & -    &- & - & - & - & - & - & -
& - \\
     & 62. & 0.18 & 0.27 & 0.2340 & 0.1 & 14.0 & 4. & 0.7 & 0.02 & 3. &
1. & 1.2 & 0.04 & 2. \\
     & 67. & 0.26 & 0.33 & 0.8626 & 0.8 & 12.7 & 3. & 0.2 & 0.05 & 2. &
30. & 0.6 & 0.17 & 1. \\
     & 73. & 0.36 & 0.41 & 1.4304 & 1.8 & 13.2 & 2. & 0.3 & 0.12 & 1. &
60. & 1.1 & 0.46 & 1. \\
     & 495.& 0.76 & 2.05 &  unst.  & -   & -    &- & - & - & - & - & - & -
& - \\ \hline
\end{tabular}
}

\newpage
\small{
\begin{tabular}{|l|l|l||l|l|l|}
\hline
n$_{H t}$ & A$_0$ & W' & n$_{H t}$ & A$_0$ & W'
\\ \hline
   1.55   &        4363  &     912&
   0.36   &        4731  &     1103\\
   1.56   &       1510   &    321&
   0.37   &       648    &    153\\
   1.57   &       811    &    175&
   0.38   &       292    &    70\\
   1.58   &        522   &     114&
   0.39   &        371   &     43\\
   1.59   &        370   &      82&
   0.40   &        123   &     30\\
   1.60   &        279   &      63&
   0.41   &        90    &       23\\
   1.61   &        219   &      50&
   0.42   &         68   &      17\\
   1.62   &        178  &      41&
          &              &         \\
   1.63   &       148   &     35&
          &              &         \\
   1.64   &       125    &     30&
          &              &         \\
   1.65   &        108   &      26&
          &              &         \\
   1.66   &         94   &      23&
          &              &         \\
   1.67   &         83   &      20&
          &              &         \\
   1.68   &         73   &      18&
          &              &         \\
   1.69   &         66   &      17&
          &              &         \\
   1.70   &         59   &      15&
          &              &         \\
\hline
\end{tabular}
}

\newpage
\textwidth 6.0in
\hoffset=-0.44in


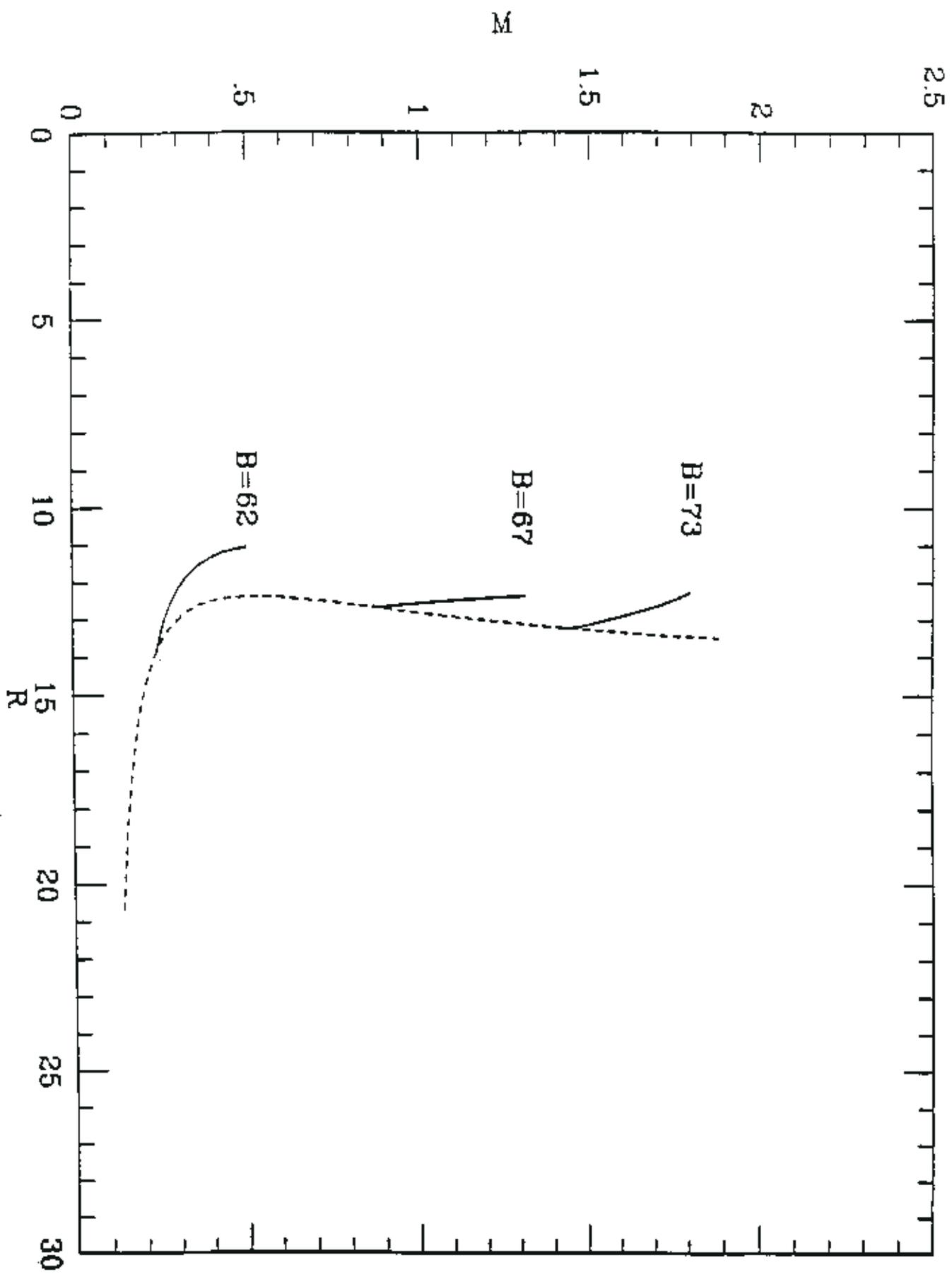
\end{document}